\documentclass[11pt,a4paper]{article}
\usepackage[utf8]{inputenc}
\usepackage[final]{acl}

\usepackage{times}
\usepackage{latexsym}

\usepackage[T1]{fontenc}

\usepackage[utf8]{inputenc}
\usepackage{CJKutf8}
\usepackage{microtype}

\usepackage{inconsolata}
\usepackage{amssymb}
\usepackage{amsmath} 
\usepackage{graphicx}
\usepackage{booktabs}
\usepackage{multirow}
\usepackage[table]{xcolor}
\usepackage{enumitem}
\usepackage[ruled,vlined]{algorithm2e}
\usepackage{amsmath}
\usepackage{minted}
\usepackage{tcolorbox}
\tcbuselibrary{minted,breakable}
\usepackage{listings}
\usepackage{xcolor}
\usepackage{subcaption}
\usepackage{hyperref}
\definecolor{codebackground}{RGB}{240, 248, 255} 
\definecolor{codecomment}{RGB}{0, 128, 0}        
\definecolor{codekeyword}{RGB}{0, 0, 255}        
\definecolor{codestring}{rgb}{0.58,0,0.82}       
\lstdefinestyle{mystyle}{
    backgroundcolor=\color{codebackground},
    commentstyle=\color{codecomment}\itshape,
    keywordstyle=\color{codekeyword}\bfseries,
    numberstyle=\tiny\color{gray},
    stringstyle=\color{codestring},
    basicstyle=\ttfamily\footnotesize,
    breakatwhitespace=false,
    breaklines=true,
    captionpos=b,
    keepspaces=true,
    numbers=left,
    numbersep=8pt,
    showspaces=false,
    showstringspaces=false,
    showtabs=false,
    tabsize=2,
    framesep=5pt
}

\title{WhispSynth: Scaling Multilingual Whisper Corpus through Real Data Curation and A Novel Pitch-free Generative Framework}

\author{
 \textbf{Tianyi Tan\textsuperscript{1,*}},
 \textbf{Jiaxin Ye\textsuperscript{2,*}},
 \textbf{Yuanming Zhang\textsuperscript{1,*}},
 \textbf{Xiaohuai Le\textsuperscript{3}},
\\
 \textbf{Xianjun Xia\textsuperscript{3}},
 \textbf{Chuanzeng Huang\textsuperscript{3}},
 \textbf{Jing Lu\textsuperscript{1}},
\\
 \textsuperscript{1}Key Laboratory of Modern Acoustics, Nanjing University, Nanjing 210093, China, 
 \\
 \textsuperscript{2}Fudan University, Shanghai, China, 
 \textsuperscript{3}ByteDance, China
 \thanks{These authors contributed equally to this work.}
}

\begin{document}
\maketitle
\begin{abstract}

Whisper generation is constrained by the difficulty of data collection. Because whispered speech has low acoustic amplitude, high-fidelity recording is challenging.
In this paper, we introduce \textbf{WhispSynth}, a large-scale multilingual corpus constructed via a novel high-fidelity generative framework. Specifically, we propose a pipeline integrating Differentiable Digital Signal Processing (DDSP)-based pitch-free method with Text-to-Speech (TTS) models. This framework refines a comprehensive collection of resources, including our newly constructed WhispNJU dataset, into 118 hours of high-fidelity whispered speech from 479 speakers. Unlike standard synthetic or noisy real data, our data engine faithfully preserves source vocal timbre and linguistic content while ensuring acoustic consistency, providing a robust foundation for text-to-whisper research.
Experimental results demonstrate that WhispSynth exhibits significantly higher quality than existing corpora. Moreover, our CosyWhisper, tuned with WhispSynth, achieves speech naturalness on par with ground-truth samples. 
The official implementation and related resources are available at \url{https://github.com/tan90xx/cosywhisper}.

\end{abstract}

\section{Introduction}

\begin{figure}[htbp]
  \centering
  \includegraphics[width=0.5\textwidth]{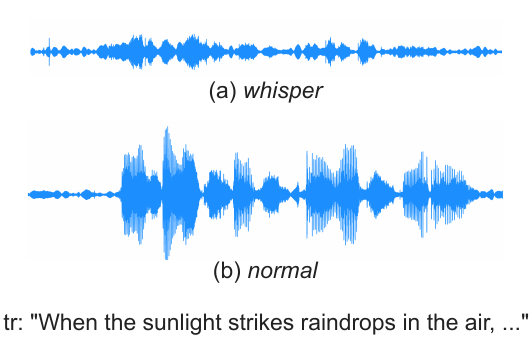}
  \caption{Different Dynamic Range. Whispers exhibit a significantly lower sound pressure level compared to normal speech, even when the linguistic content is identical.}
  \label{fig:test1}
\end{figure}

Whisper represents an innate and ubiquitous mode of human vocalisation. Primarily employed to attenuate acoustic propagation for secure communication~\cite{tartter1989s,andersen2015now, rekimoto2023wesper, hiraki2025silentwhisper}, this modality also carries significant paralinguistic weight, often signalling intimacy or inducing a soothing effect. Notably, these affective characteristics have driven the recent surge in Autonomous Sensory Meridian Response (ASMR) content across social media.

Research interest in whisper generation is rapidly growing, yet a critical data bottleneck fundamentally constrains progress. Publicly available whispered corpora are typically characterized by limited scale, heterogeneous recording quality, and a lack of speaker diversity. Specifically, the inherently low sound pressure level of whispers (see Figure~\ref{fig:test1}) makes high-fidelity capture exceptionally difficult. Furthermore, existing datasets often contain substantial acoustic artifacts that are resistant to standard speech enhancement, because whispered speech is produced with predominantly noise-driven excitation.

While Text-to-Speech (TTS) remains the primary paradigm for synthesising whisper, and general TTS has made remarkable strides in capturing nuanced prosodic features such as rhythm, intonation, and even non-verbal vocalizations~\cite{cosyvoice3, indextts2, f5tts}, achieving high-fidelity whisper generation remains a significant challenge. 
Recently, we have observed that commercial services such as Doubao AI Chatbot~\cite{doubao}, ElevenLabs TTS~\cite{eleven}, and MiniMax TTS~\cite{eleven} can produce natural-sounding whispered speech. However, these systems rely on proprietary datasets and closed-source pipelines. Consequently, the lack of high-quality, open-access resources remains a critical bottleneck that directly impedes advances in general whisper synthesis. 

To further investigate the limitations of current open-source TTS models in whisper generation, we reveal that TTS models remain severely flawed at zero-shot whisper generation. Even when provided with explicit whisper prompts or textual instructions, these models fail to bridge the domain gap, highlighting a significant deficiency in serving this special modality. 
This fragility can be attributed to two complementary biases: (\textit{i}) corpora and vocoders are predominantly optimized for modal phonation, establishing implicit priors that favor harmonic-rich signals with clear fundamental frequencies (F0); and (\textit{ii}) speaker-embedding extractors trained on such data frequently misconstrue whisper prompts as hoarseness or environmental noise rather than authentic breathy voice. 

Paradoxically, these TTS models trained on normal speech possess a latent capacity for whisper generation, derived from sporadic devoiced segments that share similar acoustic energy distributions. The overlap is anatomically grounded: whispered production suppresses vocal-fold vibration while retaining turbulent airflow at the same supra-laryngeal constriction sites exploited for voiceless obstruent in normal speech~\cite{analysis}. Consequently, the acoustic realisations of whisper and devoiced segments are governed by nearly identical aerodynamic–articulatory constraints, endowing standard models with a dormant yet exploitable reservoir of whisper-like priors.

To address these limitations, we introduce \textbf{WhispSynth}, a large-scale, multilingual corpus constructed via a novel high-fidelity generative framework. Specifically, we propose a whisper generation pipeline integrating Differentiable Digital Signal Processing (DDSP)-based pitch-free with TTS models. This framework refines a comprehensive collection of resources into 118 hours of high-fidelity 24 kHz whispers over 479 speakers. The main contributions are summarized as follows:

\begin{enumerate}
    \item \textbf{WhispReal:} We construct the first union of realistic whisper by consolidating 6 publicly available corpora and releasing our own newly recorded Mandarin dataset, \textbf{WhispNJU} ($\sim$45 h). The resulting collection, termed \textbf{WhispReal} ($\sim$118 h), provides standardized splits and metadata, establishing a robust data foundation for the field.
    
    \item \textbf{WhispSynth:} We propose a scalable data engine that synergizes state-of-the-art TTS (e.g., CosyVoice3) with a novel pitch-free method. This pipeline effectively disentangles pitch from wav signal, transforming the noisy WhispReal source into \textbf{WhispSynth}---a studio-grade synthetic corpus of approximately 118 hours. The generated whisper preserves linguistic content, reducing CER and WER by 11\%, and also improves audio quality, increasing DNSMOS by 3\%.
    
    \item \textbf{CosyWhisper:} We fine-tune CosyVoice3 on our WhispSynth to develop \textbf{CosyWhisper}. Experimental results demonstrate that training on WhispSynth yields significantly better performance than training on realistic recordings, achieving speech naturalness on par with ground-truth samples. We will release the fine-tuned model as the first open-source and reproducible \textit{text-to-whisper} system.
\end{enumerate}

\begin{table*}[!ht]
  \centering
  \footnotesize
  \resizebox{\textwidth}{!}{
    \begin{tabular}{l l l l l l l l l l}
      \toprule
      \textbf{Dataset} & \textbf{Source} & \textbf{Sample Rate} & \textbf{Size (\textit{h})} & \textbf{Avg (\textit{s})} & \textbf{Language} & \textbf{Speaker} & \textbf{P?} & \textbf{CC Type} \\
      \midrule
      UTVE-I & \cite{utve1} & 44100 & <1 & N.A. & en & 12 & N & N.A. \\
      UTVE-II & \cite{utve2} & 44100 & 1 & N.A. & en & 112 & N & N.A. \\
      AVWD & \cite{zhou2019audio} & 44100 & <2.44 & N.A. & zh & 10 & Y & N.A. \\
      Whi-spe & \cite{whispe} & 22050 & <5 & N.A. & sr & 10 & Y & N.A. \\
      AV-Whisper & \cite{tran2013audiovisual} & 48000 & <10 & N.A. & en & 11 & Y & N.A. \\
      CIAIR & \cite{ciair} & 16000 & 15 & N.A. & ja & 123 & Y & N.A. \\
      iWhisper-Mandarin & \cite{iwhisper} & 16000 & 15 & N.A. & zh & 80 & Y & N.A. \\
      wSPIRE & \cite{wspire} & 44100 & 18 & N.A. & en & 88 & Y & N.A. \\
      \rowcolor{gray!10} AISHELL6-Whisper & \cite{li2025aishell6} & 48000 & 29.75 & 9.01 & zh & 167 & Y & BY-NC-SA \\
      \rowcolor{gray!10} CHAINs (subset) & \cite{chains2006} & 44100 & 2.55 & 4.12 & en & 36 & Y & BY-ND \\
      \rowcolor{gray!10} EARs (subset) & \cite{richter2024ears} & 48000 & 3.22 & 16.58 & en & 107 & Y & BY-NC \\
      \rowcolor{gray!10} Expresso (subset) & \cite{nguyen2023expresso} & 48000 & 1.32 & 3.13 & en & 4 & Y & BY-NC \\
      \rowcolor{gray!10} Whisper40 & \cite{yang2024whisper40} & 16000 & 6.10 & 12.25 & zh & 40 & Y & N.A. \\
      \rowcolor{gray!10} wTIMIT & \cite{li2005establishment} & 44100 & 29.44 & 5.04 & en & 48 & Y & N.A. \\
      \specialrule{0em}{1.pt}{0.6pt}
        \hline
        \specialrule{0em}{1.pt}{1.0pt}
      \rowcolor{gray!10} WhispNJU & Our new recording & 44100 & 45.24 & 9.01 & zh & 77 & Y & BY-NC \\
      WhispReal & Our new curation & Mixed & 117.62 & 9.91 & en, zh & 479 & Y & Mixed \\
      WhispSynth & Synthesized from WhispReal & 24000 & $\approx$118 & $\approx$10 & en, zh & 479 & N & MIT \\
      \bottomrule
    \end{tabular}
  }
  \caption{Overview of datasets used in this work, with the highlighted (gray) rows indicating all obtainable real whisper corpora included in our study (up to 2025-01-05). Entries include data duration (Size), average utterance length (Avg), language, speaker count, whether the paired normal segments are available (P?), and license type (CC Type), ``Mixed'' indicates multiple license terms or sample rate, ``N.A.'' denotes unknown or unspecified.}
  \label{tab:test1}
\end{table*}

\section{WhispReal}

WhispReal dataset is a curated collection of existing publicly available whispered speech corpora combined with our proprietary \textbf{WhispNJU} dataset. As summarized in Table~\ref{tab:test1}, the number of open-source whisper datasets is limited, and they exhibit significant variations in scale, linguistic diversity, and accessibility. The licensing terms (CC Type) for these datasets are also heterogeneous, including Creative Commons Attribution-NonCommercial-ShareAlike (NC-SA), Attribution-NoDerivatives (ND), Attribution-NonCommercial (NC), among others, while the license information for some datasets remains unspecified (N.A.). This heterogeneity directly impacts data availability, redistribution rights, and the freedom to utilize the data for academic and commercial research. In constructing WhispReal, we meticulously respected and adhered to the licensing agreements of all source datasets, ensuring the full legal compliance of the resulting collection.

Existing whispered speech corpora documented in the literature can be broadly categorized into three primary types, with a subset being publicly available as highlighted in gray in Table~\ref{tab:test1}:

\noindent \textbf{Pure Whisper Datasets} These collections are specifically dedicated to whispered vocalizations, designed for targeted acoustic analysis~\cite{li2005establishment, jou2005whispery, utve1, lim2011computational, utve2, lee2014whispered, wspire, hiraki2022silentwhisper, yang2024whisper40}. Among these, wTIMIT~\cite{li2005establishment} is the most notable and widely adopted. It is a classic, relatively large‑scale English whispered speech corpus that also includes parallel normal speech. The linguistic content of these collections is largely confined to the word or phrase level.

\noindent \textbf{Subsets within Expressive Speech Databases} Several speech corpora designed for expressive analysis incorporate whispered speech as one vocalization style~\cite{chains2006, nguyen2023expresso, richter2024ears}. These datasets often contain rich emotional annotations~\cite{DBLP:conf/acl/MaZYLGZ024, DBLP:conf/interspeech/MaCZZCLY0H24,DBLP:conf/mm/YeWWMH0S23, DBLP:conf/icassp/YeWWXLS23,demoface:conf/icml/YeCS25}, facilitating research on affective whispering, while the whisper portions themselves are typically limited in size.

\noindent \textbf{Audiovisual Whisper Collections} This is an emerging category that integrates whispered speech with synchronized visual recordings~\cite{ciair, tran2013audiovisual, petridis2018visual, zhou2019audio, li2025aishell6}. Within this domain, AISHELL6-Whisper~\cite{li2025aishell6} represents one of the largest Chinese datasets, characterized by a substantial number of speakers, which makes it well-suited for multimodal analysis and robustness research. It is worth noting that the upcoming dataset, WhispNJU, also falls into this category. However, due to the heightened complexity of the recording process, the scalability of such audiovisual collections is constrained.

In the following sections, we present information on the existing available whispered data sources and a unique processing required for each. This is followed by a comprehensive account of the data collection procedures for our dataset.

\subsection{Existing Available Resources}\label{sec:data_split}

\textbf{AISHELL6-Whisper}~\cite{li2025aishell6}: it provides audio-visual whisper speech dataset, featuring 30 hours each of whisper speech and parallel normal speech, with synchronized frontal facial videos. We used only the audio and followed the train/test/valid splits. It should be noted that some utterances within the retained set still lack paired counterparts.

\noindent \textbf{wTIMIT}~\cite{lim2011computational}: it is constructed after the TIMIT dataset~\cite{garofolo1993darpa}. The first phase of the project collected the speech from 20 Singaporeans, and the second phase collected the speech from 28 North Americans. After excluding files with evident sampling errors, we retained all remaining audio files. It should be noted that some utterances within the retained set still lack whispered counterparts. Subsequently, we reorganized the dataset by speaker into a 60\%–20\%–20\% split for training, validation, and testing, respectively. 

\noindent \textbf{Whisper40}~\cite{yang2024whisper40}: it is a Mandarin Chinese corpus comprising whispered and normal speech recordings from 40 speakers without corresponding transcripts. However, since the normal and whispered speech pairs share identical linguistic content, the normal speech is clearly articulated, and the text is consistent across speakers, the authors performed manual transcription and verification for all utterances. We followed the train/test/valid splits of the original dataset.

\noindent \textbf{CHAINS}~\cite{chains2006}: \textbf{CHA}racterizing \textbf{IN}dividual \textbf{S}peakers contains recordings from 36 speakers (8 from the UK/US and 28 from Ireland), each producing 37 utterances across six different speaking conditions: solo speech, retelling, synchronous speech, repetitive synchronous imitation, fast speech, and whispered speech. We extracted the solo speech and whispered speech from CHAINs and segmented its four short fables (Cinderella, Rainbow text, North Wind and the Sun, and Members of the Body) into sentences to ensure that each audio file is under 30 seconds in duration, and split the data by speaker into 60\% training, 20\% validation, and 20\% test sets.

\noindent \textbf{Expresso}~\cite{nguyen2023expresso}: it is a high-quality (48~kHz) expressive speech dataset containing expressive read speech in 8 styles and improvised dialogues in 26 styles, recorded from 4 speakers. We selected default style and whisper style of it. Following a consistent partitioning scheme, we split the data by speaker into 60\% training, 20\% validation, and 20\% test sets.

\noindent \textbf{EARs}~\cite{richter2024ears} : \textbf{E}xpressive \textbf{A}nechoic \textbf{R}ecordings of \textbf{S}peech is a high‑quality expressive speech collection, recorded at 48~kHz and comprising 107 speakers from diverse backgrounds, totaling approximately 100 hours of clean, anechoic speech. EARs covers the full dynamic range of vocal expression, from whispering to yelling and screaming. For this study, we extracted the whispered and regular speech subsets (split: 60\% train / 20\% valid / 20\% test by speaker).

\subsection{The WhispNJU corpus}
The WhispNJU corpus consists of 85 hours of paired whispered and normal speech from 77 speakers of students recruited at X University. WhispNJU is designed to facilitate research and development in Mandarin whispered ASR. It is constructed after the THCHS-30 dataset~\cite{wang2015thchs}, which is usually used for the study of Mandarin speech recognition. 

\noindent \textbf{Speaker Recruitment}
The dataset was recorded by 77 native Mandarin speakers (37 male and 40 female), aged 22-26, recruited from X University. All participants met the following selection criteria essential for speech production research:
\begin{itemize}[noitemsep, topsep=0pt, partopsep=0pt]
    \item Native proficiency in Mandarin Chinese.
    \item No diagnosed speech or hearing disorders.
    \item No acute oral/auditory health conditions.
\end{itemize}
These criteria ensured consistent recording quality while complying with ethical research standards. Participants received compensation and provided informed consent.

\noindent \textbf{Record Settings}
The participants were instructed to produce each sentence twice: once in normal speech and once in whispered speech. Due to the time-consuming nature of the recordings, sessions could be divided into multiple sittings, and submissions were only accepted after all recordings were completed. To ensure universal applicability, no restrictions were placed on the speakers’ regional origin or accent. During recording, no obstructions were allowed between the speaker and the recording device to avoid interfering with airflow pickup by the microphone.

\noindent \textbf{Data preparation}
Following the partitioning scheme of the THCHS-30 dataset~\cite{wang2015thchs}, our dataset is divided into four groups according to the recording text: A  (sentence ID from 1 to 250), B (sentence ID from 251 to 500), C (sentence ID form 501 to 750), and D(sentence ID from 751 to 1000). Each participant recorded 250 sentences (approximately 50 words per sentence), totaling approximately one hour of recording per speaker. Speakers were randomly assigned unique identification numbers and evenly distributed across the four groups.

\section{WhispSynth}
\subsection{Generation Pipeline}
Although recent advances in TTS synthesis have enabled the generation of highly natural and intelligible speech, when synthesizing whispered speech many state-of-the-art systems tend to produce residual pitch or fundamental frequency (F0) components. To address this issue, we propose a post-processing pipeline that detects and removes pitch components from synthesized whispers while preserving the desired noise-like whisper quality. Figure~\ref{fig:test2} outlines the proposed procedure.

The core idea is to leverage a DDSP decomposition to isolate and suppress harmonic content in pitch-contaminated segments, followed by an overlap-add (OLA) reconstruction to ensure smooth transitions. The full algorithm is provided in Appendix~\ref{app:algorithm}.

\begin{figure}[htbp]
  \centering
  \includegraphics[width=0.5\textwidth]{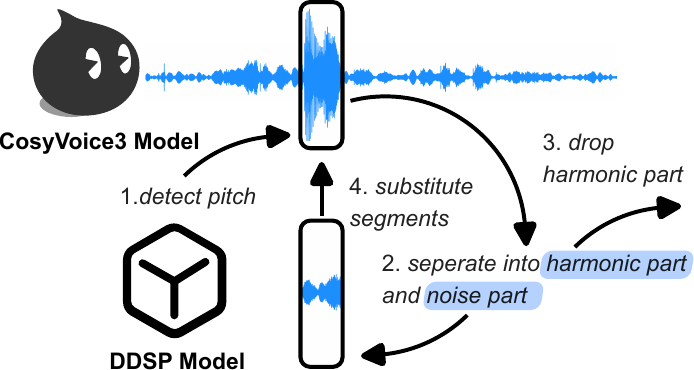}
  \caption{Visualization of the WhispSynth's generation pipeline. We apply CosyVoice3 and DDSP Model to generate pitch-free voice.}
  \label{fig:test2}
\end{figure}

\subsection{CosyVoice3: State-of-the-Art Speech Synthesis}\label{sec:cosy}
CosyVoice3~\cite{cosyvoice3}, developed by Alibaba Group, represents a next-generation speech synthesis model designed for high-quality and natural speech in open-environment scenarios. It integrates a multi-task speech tokenizer, scalable Differentiable Reward Optimization for post-training, and is trained on an ultra-large-scale multilingual corpus with instruction-following controllability. 

The model exhibits particular strength in whisper synthesis. Its multi-task tokenizer, trained jointly using speech emotion recognition and audio event detection tasks, effectively encodes paralinguistic features such as softness and breathiness. Moreover, its million-hour-scale training corpus, drawn from diverse real-world recordings, contains abundant whisper and low-volume speech samples, enabling robust learning of such vocal qualities. Nonetheless, during whisper synthesis, the model occasionally exhibits fundamental frequency artifacts. This issue stems from the fact that the neural modules are predominantly trained on phonated speech, leading them to inadvertently introduce subtle periodicity even when generating whisper-like outputs.

\subsection{Pitch-free with DDSP: An Overview}
The DDSP model~\cite{engel2020ddsp}, proposed by researchers from Google, adopts the harmonic-plus-noise model~\cite{serra1990spectral} to decompose a monophonic sound into a harmonic component ${\mathrm{H}}$ and a noise component ${\mathrm{N}}$:
\begin{equation}
\mathbf{S}[i] = \mathbf{H}[i]+\mathbf{N}[i].
\end{equation}
Given the upsampled estimated $\widetilde{F}_{0}[i]$, in this work we approximate ${\mathbf{H}}$ by a sawtooth signal~\cite{wu2022ddsp}, which contains an equal number of even and odd harmonics with decaying magnitudes:
\begin{equation}
\widetilde{\mathbf{H}}[i] = \sum_{k=1}^K \frac{1}{k} \sin(\phi_k[i]).
\end{equation}
The instantaneous phase $\phi_k(t)$ is obtained by cumulative summation of the instantaneous frequency samples $k\widetilde{F}_{0}[i]$:
\begin{equation}
\phi_k[i]=2\pi \sum_{n=0}^{N}k \widetilde{F}_{0}[i],
\end{equation}
where $\widetilde{\mathbf{H}}$ is treated as the “excitation signal” and shaped into the desirable $\mathbf{y}_\mathrm{h}$ by means of a linear time-varying finite impulse response (LTV-FIR) filter $\psi_{\mathrm{h}}(i) \in \mathbb{R}^{L_{\mathrm{h}}}$:
\begin{equation}
\bar{\mathbf{H}}[i] = \widetilde{\mathbf{H}}[i] \ast \psi_{\mathrm{h}}[i],
\end{equation}
where $\mathbf{N}$ is approximated by convolving a uniform distributed noise signal $\zeta$ ranging from $-$1 to 1 (with the same length as a frame) with an LTV-FIR filter $\psi_{\mathrm{n}}[i] \in \mathbb{R}^{L_{\mathrm{n}}}$ estimated per frame:
\begin{equation}
\bar{\mathbf{N}}[i] = \zeta[i] \ast \psi_{\mathrm{n}}[i].
\label{noise_filter}
\end{equation}
Jointly, the parameters $\Phi := \{\widetilde{\mathrm{F_0}}[i],\psi_{\mathrm{h}}[i],\psi_{\mathrm{n}}[i]\}_{i=1}^{N}$ are estimated from an input mel-spectrogram $\mathbf{X}$ per frame by an NN mapping function $f_{\mathrm{NN}}$~\cite{gulati2020conformer}:
\begin{equation}
\Phi = f_{\mathrm{NN}}(\mathbf{X}).
\end{equation}

DDSP entails a source-filter model~\cite{wang2019neural} that can successfully separate the noise component from normal speech.

\subsection{Pitch-free with DDSP: Training Pipeline}

Based on the prior work of DDSP in singing vocoders~\cite{wu2022ddsp}, the authors identified a critical limitation: the model produces buzzing artifacts in unvoiced and semi-voiced segments when used as a subtractive synthesizer\footnote{\small{\url{https://github.com/YatingMusic/ddsp-singing-vocoders/tree/main/postprocessing}}}. To overcome this limitation and adapt DDSP for high‑quality whispered speech synthesis, we propose the following three training strategies:

\noindent (i) \textbf{Adversarial Training with Normal Speech:}
We integrate the DDSP generator into the BigVGAN~\cite{lee2022bigvgan} framework, applying a multi‑resolution discriminator (MRD) and a multi‑period discriminator (SPD) with least‑square GAN loss $\mathcal{L}_{\mathrm{adv}}$~\cite{mao2017least}. This adversarial training is essential because GAN‑based discrimination has been shown to effectively improve the modeling of harmonic structures and suppress artifact‑prone synthesis patterns.

\noindent (ii) \textbf{Continued Training with Whispered Speech:}
After initially training the DDSP vocoder on normal speech data, we continue training it on the WhispReal dataset without adversarial objectives for faster training speed. This two‑stage procedure allows the model to first learn robust pitch‑conditioned synthesis, then specialize in whispered speech generation while avoiding instability that might arise from direct training on limited whispered data.

\noindent (iii) \textbf{Semi‑supervised Dual‑focus Training:}
We observe that the WhispReal dataset contains samples with perceptible pitch contours resembling normal speech, due to speaker errors or temporary vocal fatigue. Therefore the training scheme always treats the output as a sum of harmonic and stochastic components, but shifts the learning emphasis depending on the input type: whispered/normal speech share loss, but the gradient re-weights harmonic/noise terms so each half teaches the other. This dual‑focus approach enables mutually reinforcing improvements in both harmonic and stochastic modeling.

\section{CosyWhisper}

We fine-tune CosyVoice3 with Whisper, following the exact procedure outlined in the official script. The CosyVoice3 architecture comprises three primary components: the text-to-speech large language model (LLM), the Conditional Flow Matching (CFM) model, and HiFi-GAN vocoder. In our scenario, the distinction between whispered and normal speech lies primarily in acoustic modeling rather than semantic content. Therefore, we selectively fine-tune only the CFM model, which is responsible for converting semantic tokens into high-fidelity acoustic features, while keeping the LM and HiFi-GAN unchanged.

Although the official script currently supports only LLM training (line 65: “Run train. We only support LLM training for now”), we successfully extended it to CFM fine-tuning. Key adaptations include adding a token projection layer, replacing the encoder with a direct embedding lookup, and revising the conditioning mechanism. These changes improve token-acoustic alignment and model efficiency. Implementation specifics are provided in the Appendix~\ref{app:code}.
We refer to the resulting fine-tuned model as \textbf{CosyWhisper}, a name that reflects its origin in the CosyVoice3 architecture and its specialized capability in Whisper-based speech synthesis.

\section{Experiments}
\subsection{Experimental setups}
\textbf{Data Settings} We used the English and Mandarin Chinese subsets of the WhispReal dataset for training and evaluation. And the WhispSynth dataset is partitioned accordingly. For each source, data were divided as described in Section~\ref{sec:data_split}. The validation set was used for model checkpoint selection, while the final sound quality evaluation was performed only on the test set. Detailed statistics of the dataset distribution are provided in Table~\ref{tab:test2}.

\begin{table}[htbp]
  \centering
  \footnotesize
  \resizebox{0.5\textwidth}{!}{ 
    \begin{tabular}{l c c c c }
    \toprule
        \multirow{2}{*}{~} & \multicolumn{2}{c}{\textbf{en}} & \multicolumn{2}{c}{\textbf{zh}} \\
        \specialrule{0em}{1.pt}{0.6pt}
        \cline{2-5}
        \specialrule{0em}{1.pt}{0.6pt}
        ~ & \textbf{\# Train} & \textbf{\# Test} & \textbf{\# Train} & \textbf{\# Test} \\
        \midrule
        File Count  & 20694 & 4770 & 32704 & 7256 \\
        Size (h) & 29.62 & 6.91 & 65.86 & 15.25 \\
        Avg (s)& 5.15 & 5.21 & 7.25 & 7.57 \\ 
        Speaker Count & 78M 87F & 15M 15F & 206M 194F & 26M 25F \\ 
    \bottomrule
    \end{tabular}
    }
    \caption{The statistics of the WhispReal dataset train/test distribution.}
    \label{tab:test2}
\end{table}
\begin{table*}[htbp]
\footnotesize
  \centering
  \resizebox{0.85\textwidth}{!}
  {
  \begin{tabular}{llccccc}
    \toprule
    \multirow{2}{*}{\textbf{Dataset}} & \multirow{2}{*}{\textbf{Source}} & \multirow{2}{*}{\textbf{Language}} & \multicolumn{2}{c}{\textbf{Naturalness} $\uparrow$ } & \textbf{Intelligibility} $\downarrow$ & \textbf{F0} $\downarrow$\\ 
    \specialrule{0em}{1.pt}{0.6pt}
    \cline{4-7}
    \specialrule{0em}{1.pt}{1.0pt}
    {} & {} & {} & {\textbf{DNSMOS}} & {\textbf{UTMOS}} & {\textbf{CER} / \textbf{WER} (\%)} & {\textbf{VTR(\%)}}\\
    \midrule
    \specialrule{0em}{1.pt}{0.6pt}
    wTIMIT & \cite{li2005establishment} & en  & 2.76 & 1.31 & 50.99 & 0.69\\
    CHAINs (\textit{subset})  & \cite{chains2006}  & en  & 2.76 & 1.48 & 11.55 & 0.95\\
    Expresso (\textit{subset}) &  \cite{nguyen2023expresso} & en  & 3.24 & 1.50 &  7.77 & 0.58\\
    EARs (\textit{subset})   &  \cite{richter2024ears}  & en  & 3.43 & 2.01 &  6.31 & 0.30\\
    Whisper40     & \cite{yang2024whisper40}     & zh  & 2.61 & 1.27 & 72.80 & 0.99\\
    AISHELL6-Whisper  & \cite{li2025aishell6} & zh & 2.75 & 1.68 & 38.26 & 0.94\\
    WhispNJU      & Ours     & zh  & 2.90 & 1.33 & 36.74 & 0.98\\
    \specialrule{0em}{1.pt}{0.6pt}
    \midrule
    \specialrule{0em}{1.pt}{1.0pt}
    WhispReal & Ours  & en,zh  & {2.80} & {1.44} & {39.30/37.58} & 0.88\\
    WhispSynth& Ours  & en,zh  & \textbf{2.89} & \textbf{1.46} & \textbf{31.16/20.98} & 0.87\\ 
    \bottomrule
  \end{tabular}}
  \caption{Objective evaluation results on open-sourced and our proposed datasets. Comprehensive evaluation on whisper speech conversion and synthesis. \textbf{Quality}: Subjective naturalness (DNSMOS/UTMOS $\uparrow$). \textbf{Intelligibility}: Character/Word Error Rate ($\downarrow$). \textbf{Pitch}: Voiced Time Ratio (VTR) ($\downarrow$).}
  \label{tab:evaluation}
\end{table*}
\vspace{-2mm}
\noindent \textbf{Training Settings}
We transformed the normalized speech in all the datasets into mel-spectrograms with a frame length of 1280 and a hop length of 320. Adversarial training was conducted using a batch size of 8, while standard (non-adversarial) training employed a batch size of 32. All experiments were performed on eight V100-32GB GPUs.

\noindent \textbf{Evaluation Settings} 
The generation performance is evaluated using both subjective and objective metrics. 
In terms of subjective evaluation, the proposed and baseline systems were evaluated using a 5-point Whisper-likeness Mean Opinion Score (W-MOS). It measured the perceptual similarity between the synthesized whisper and the listener’s mental prototype of a whisper, ranging from 0 (“Totally Not similar”) to 5 (“Extremely similar”). We have developed an automated subjective evaluation interface, as shown in Appendix~\ref{app:subjective}.
For objective assessment, we adopt standard TTS metrics, as no widely used metrics are specifically designed for whisper.
For the naturalness, we measure spectral differences with Mel Cepstral Distortion (MCD)~\cite{visualvoicecloning/ChenTQZLW22}, DNSMOS~\cite{dnsmos:conf/icassp/ReddyGC21}, and UTMOS~\cite{utmos} to estimate perceptual audio quality. We likewise calculate the Word Error Rate (WER)~\cite{WER/WangSZRBSXJRS18} for English and Character Error Rate (CER)~\cite{CER/xu2025fireredasr} for Chinese to gauge intelligibility using SOTA multilingual ASR model \texttt{Fun-ASR-nano-2512}~\cite{an2025fun}. 
For the timbre, we calculate cosine similarity metrics based on \texttt{ECAPA-TDNN}~\cite{DBLP:conf/interspeech/DesplanquesTD20} to obtain speaker identity similarity (SpkSim). 
Additionally, we evaluate Voiced Time Ratio (VTR) for pitch-free assessment.

\subsection{Evaluating on Benchmarks}
Based on the objective evaluation results in Table~\ref{tab:evaluation}, several key observations can be made regarding the quality of whispered corpora. Among the publicly available English datasets, the EARs subset demonstrates superior performance, achieving the highest naturalness scores (DNSMOS: 3.43, UTMOS: 2.01) and the lowest CER: 6.31\%, indicating its high-quality whisper characteristics. For Chinese datasets, our proposed WhispNJU corpus shows a favorable balance, attaining a higher DNSMOS score (2.90) compared to other Chinese whisper collections like Whisper40 (2.61) and AISHELL6-Whisper (2.75). The primary advantages of our newly introduced datasets, WhispReal and WhispSynth, are threefold. First, they provide multilingual coverage for both English and Chinese, addressing a gap in existing resources that are predominantly monolingual. Second, WhispSynth exhibits exceptional intelligibility, achieving the lowest error rates (CER: 31.16\%, WER: 20.98\%) among all evaluated datasets, which is beneficial for speech recognition and synthesis tasks requiring high clarity. Third, it maintains a balanced naturalness, with competitive DNSMOS (2.89) and UTMOS (1.46) scores while demonstrating a reasonable pitch profile (VTR: 0.87) comparable to high-quality datasets like Expresso. 

Overall, WhispSynth presents an optimal trade-off between intelligibility and naturalness, making it a suitable resource for whisper-based speech processing applications. In addition, the results also reveal that existing objective metrics are poorly aligned with whisper characteristics; ultimately, human judgment remains indispensable.

\subsection{Experiments for generated samples}

\subsubsection{Comparative Systems} 
\noindent \textbf{Whisper-Effect}~\cite{whisper_effect2025} is a designed acoustic perturbation that simulates whisper by applying a high-pass filter combined with white noise.

\noindent \textbf{toWhisper}~\cite{toWhisper2017} is an Linear Predictive Coding (LPC)-based tool that synthesizes whispered speech by processing white noise through a vocal tract filter estimated via LPC vocoding.

\noindent \textbf{Normal2Whisper}~\cite{normal2whisper2023} converts normal speech to whisper through a two-stage process (Glottal source removal and Spectral modification) by the WORLD vocoder~\cite{morise2016world}.

\noindent \textbf{SeedVC}~\cite{liu2024zero} is a zero-shot voice conversion framework that disentangles representations of content and speaker information.

\noindent \textbf{CosyVoice3}~\cite{cosyvoice3} is a multilingual TTS model for high-quality and efficient generation. 

\begin{table*}[!ht]
  \centering
  \footnotesize
  \setlength{\tabcolsep}{4.5pt} 
  \resizebox{\textwidth}{!}{
    \begin{tabular}{l l c c c c c c c}
    \toprule
        \multirow{2}{*}{\textbf{Method}} & 
        \multirow{2}{*}{\textbf{Source}} & 
        \textbf{Whisperiness} &
        \multicolumn{3}{c}{\textbf{Naturalness}} & 
        \textbf{Intelligibility} & 
        \textbf{Timbre} & 
        \textbf{Pitch} \\
        \cmidrule(lr){3-9}
        & & 
        \textbf{W-MOS $\uparrow$} &
        \textbf{DNSMOS $\uparrow$} & 
        \textbf{UTMOS $\uparrow$} & 
        \textbf{MCD $\downarrow$} & 
        \textbf{CER/WER (\%) $\downarrow$} & 
        \textbf{SpkSim $\uparrow$} & 
        \textbf{VTR (\%) $\downarrow$} \\
    \midrule
    Ground Truth & Test Set of WhispReal  & 4.33$\pm$0.33 & 2.80 & 1.44 & 0.00 & 39.30/37.58 & 1.00 & 0.88 \\
    \midrule
        \multicolumn{8}{l}{\textbf{\textit{Normal-to-Whisper Conversion}}} \\
        \specialrule{0em}{1.pt}{0.6pt}
        Whisper-Effect & \cite{whisper_effect2025} & - & 3.14 & 1.30 & 70.23 & 59.46/80.35 & 0.52 & 0.12 \\
        toWhisper & \cite{toWhisper2017} & 1.02$\pm$0.29 &2.94 & 1.35 & 55.73 & 12.07/~9.27 & 0.58 & 0.92 \\
        Normal2Whisper & \cite{normal2whisper2023} & - & 2.89 & 1.28 & 60.00 & 11.92/28.28 & 0.59 & 0.94 \\
        SeedVC & \multirow{1}{*}{\cite{liu2024zero}} & 2.18$\pm$0.47 & 3.01 & 1.61 & 54.85 & 58.65/19.97 & 0.66 & 0.71 \\
        \specialrule{0em}{1.pt}{0.5pt}
\rowcolor{gray!20} Pitch-free Model & Ours & 1.30$\pm$0.29 & 2.90 & 1.36 & 63.05 & 21.31/45.18 & 0.63 & 0.98 \\
    \midrule
        \multicolumn{8}{l}{\textbf{\textit{Text-to-Whisper Synthesis}}} \\ 
        \specialrule{0em}{1.pt}{0.6pt}
        CosyVoice3  & \multirow{1}{*}{\cite{cosyvoice3}} & 3.40$\pm$0.51 & 3.00 & 1.47 & 56.03 & 12.51/9.81 & 0.83 & 0.86 \\
       \specialrule{0em}{1.pt}{0.5pt}
\rowcolor{gray!20}  CosyWhisper & {Ours} & \textbf{4.53$\pm$0.20} & 3.08 & 1.48 & 59.31 & 12.76/29.22 & 0.80 & 0.88 \\
    \bottomrule
    \end{tabular}}
    \caption{Evaluation on whisper speech conversion and synthesis. \textbf{Quality}: Objective whisperiness (W-MOS $\uparrow$), Subjective naturalness (DNSMOS/UTMOS $\uparrow$) and objective distortion (MCD $\downarrow$). \textbf{Intelligibility}: Character/Word Error Rate ($\downarrow$). \textbf{Timbre}: Speaker similarity cosine similarity ($\uparrow$). \textbf{Pitch}: Voiced Time Ratio (VTR) ($\downarrow$). 
    }
    \label{tab:main_results}
\end{table*}

\subsection{Experiments for CosyWhisper}

\subsubsection{Quantitative Evaluation}

The objective results in Table~\ref{tab:main_results} highlight two key findings.  
First, in \textit{Normal-to-Whisper Conversion}, voice conversion models like SeedVC yield high speaker similarity and intelligibility, but their whisperness remains limited (W-MOS: 2.18). Although traditional baselines (e.g., toWhisper) are non-trainable, lack adaptability, and yield lower W-MOS despite decent objective scores. Our pitch-free model is trainable and better captures natural whisper variability, confirming that not fixed transformations for authentic whisper synthesis.
Second, in \textit{Text-to-Whisper Synthesis}, even strong TTS systems like CosyVoice3 produce speech that is intelligible and natural, yet still perceptibly non-whisper (W-MOS: 3.40). In contrast, our {CosyWhisper} achieves a W-MOS of 4.53---the highest among all synthesized approaches and even better than the ground-truth (4.33), demonstrating our superiority to generate authentic whisper from text.

\begin{figure}[htbp]
  \centering
  \includegraphics[width=0.5\textwidth]{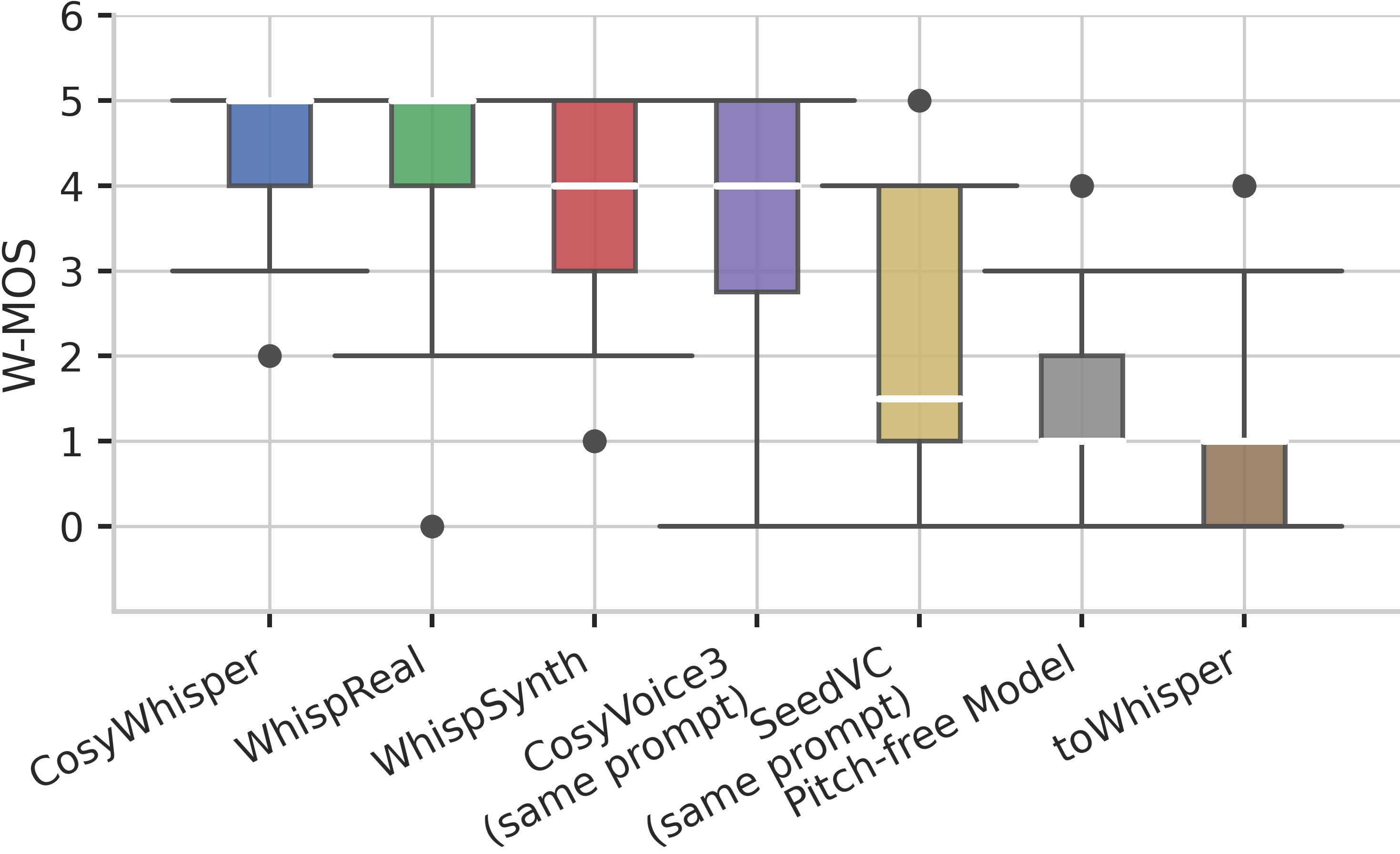}
  \caption{Results of the listening test. Whisper-likeness Mean Opinion Score (W-MOS) based
on 20 participants visualized in a standard box plot.}
  \label{fig:test3}
\end{figure}
\vspace{-2mm}

For subjective evaluation, the W-MOS box plot from the listening test statistically validates that listeners perceive the output of CosyWhisper as similar to natural whisper. The interquartile range (IQR) for CosyWhisper is expected to be compact and positioned high on the scale (near 4.0), with a median close to 4.1, indicating consistent high-quality ratings from participants with low disagreement. In contrast, the boxes for other methods (e.g., CosyVoice3, SeedVC) would be located lower, with potentially larger IQRs, reflecting greater variability in listener perception and lower consensus on their whisper authenticity. CosyWhisper successfully bridges the gap between high intelligibility and authentic whisper perceptual quality.

\subsection{Ablation on Tuning Dataset}

To evaluate the efficacy of our synthetic data for text-to-whisper synthesis, we fine-tune our CosyWhisper on two variants of the corpus: WhispReal and our synthetic WhispSynth. As shown in Table~\ref{tab:finetune}, 
WhispSynth consistently outperforms across all key metrics: it reduces CER/WER by 46\%, lowers pitch distortion (VTR) by 9\%, and improves naturalness by 8\%, despite being fully synthetic. 
This confirms that WhispSynth not only mitigates the noise and inconsistency inherent in real whisper data but also provides high-fidelity data that better supports stable whisper generation from text.

\begin{table}[h]
  \centering
  \footnotesize
  \resizebox{0.47\textwidth}{!}{ 
    \begin{tabular}{ c c c c c c }
    \toprule
        \multirow{2}{*}{\textbf{Dataset}} & \multicolumn{2}{c}{\textbf{Naturalness $\uparrow$} }& \textbf{Intelligibility $\downarrow$} & \textbf{Timbre $\uparrow$} & \textbf{F0 $\downarrow$} \\
        \specialrule{0em}{1.pt}{0.6pt}
        \cline{2-6}
        \specialrule{0em}{1.pt}{0.6pt}
        ~ & \textbf{DNSMOS} & \textbf{UTMOS} & {\textbf{CER} / \textbf{WER} (\%)} & \textbf{Cosine} & \textbf{VTR(\%)} \\
    \specialrule{0em}{1.pt}{0.6pt}
    \midrule
    \specialrule{0em}{1.pt}{0.6pt}
        \multirow{1}{*}{\textbf{\textit{WhispReal}}} & 2.93 & 1.33 & 28.3/46.5 & 0.75 & 0.77 \\
        \specialrule{0em}{1.pt}{0.6pt}
        \multirow{1}{*}{\textbf{\textit{WhispSynth}}} & \textbf{3.08} & \textbf{1.48} & \textbf{12.8/29.2} & \textbf{0.80} & \textbf{0.70} \\
    \bottomrule
    \end{tabular}}
    \caption{Performance comparison of our CosyWhisper fine-tuned on {WhispReal} versus {WhispSynth}.} 
    \label{tab:finetune}
\end{table}
\vspace{-2mm}

\section{Conclusion}

In this paper, we address the fundamental data scarcity challenge in whisper research by introducing WhispSynth, a large-scale, high-fidelity multilingual whisper corpus. Our core contribution is a novel generative framework that integrates a DDSP-based pitch-free method with advanced TTS models, transforming diverse and noisy real whispered recordings into a clean, studio-grade synthetic dataset. This pipeline ensures the faithful preservation of vocal timbre and linguistic content while significantly enhancing acoustic quality.

\newpage
\section*{Limitations}

The impact of non-linguistic variations (e.g., the use of different microphones) on model performance was not systematically assessed. Although existing research suggests that CosyVoice3 are relatively robust to such variations, the influence of hardware differences in real-world deployment scenarios on specific task performance requires further verification.
Besides, to address security concerns, the CosyWhisper model used in this study will be released with an automatically embedded real-time audio watermark. While this measure is crucial for responsible usage tracking, it may have a potential impact on the acoustic properties of the synthesized audio and subsequent analyses.

\section*{Ethics Statement}
Our paper evaluated various methods that could make developing text-to-whisper synthesize systems more viable for languages where paired whisper and transcriptions are difficult to obtain. In our experiments, we only used already publicly available data (CHAINs, EARs, Expresso, Whisper40) or data for which we have obtained informed consent for public release from the data custodians (AISHELL6-Whisper, wTIMIT). To make our findings as relevant as possible for other language projects, we minimized the amount of computing time used.

\bibliography{custom}

\clearpage
\onecolumn 
\appendix

\section*{Appendix}
\label{sec:appendix}

\section{Algorithm}\label{app:algorithm}
Algorithm~\ref{alg:whisper-synthesis} outlines the proposed procedure. 
First, an initial whisper is generated using CosyVoice3, conditioned on a prompt audio and its transcript; note that an instruction is optional. The pitch detection module in DDSP then identifies segments containing residual F0. For each such segment, DDSP decomposition separates the harmonic and noise components. The harmonic part is dropped, while the noise component, which carries the whisper’s spectrnal envelope and aperiodic excitation, is retained. Finally, an OLA operation is applied to reconstruct a continuous, pitch-free whisper waveform.

\SetAlFnt{\small}
\SetAlCapFnt{\small}
\SetAlCapNameFnt{\small}
\newcommand{\To}{\textbf{to}\ }
\begin{algorithm}[htbp]
\caption{Whisper Synthesis with Pitch-Aware Segment Replacement}
\label{alg:whisper-synthesis}
\textbf{Input:} 
    Prompt audio $\mathbf{A}_p$, 
    Corresponding text $T_p$, 
    Instruction $I$ \\
\textbf{Output:} 
    Synthesized whisper audio $\mathbf{A}_w$
\DontPrintSemicolon
Initialize $\mathbf{A}_w \gets \emptyset$ \tcp*{Final whisper audio}

$\mathbf{A}_w^{\text{init}} \gets \text{CosyVoice3}(\mathbf{A}_p, T_p, I)$ \;

$\mathcal{P} \gets \text{DDSP\_PitchDetection}(\mathbf{A}_w^{\text{init}})$ \;
\If{$\mathcal{P} \neq \emptyset$}{
    \ForEach{pitch segment $\mathbf{S}_i \in \mathcal{P}$}{
        $(\mathbf{H}_i, \mathbf{N}_i) \gets \text{DDSP\_Decompose}(\mathbf{S}_i)$ \;
        \tcp{$\mathbf{H}_i$: harmonic segment, $\mathbf{N}_i$: noise segment}
        
        $\mathbf{S}_i^{\text{noise}} \gets \mathbf{N}_i$ \;
        
        $\mathbf{S}_i^{\text{replaced}} \gets \text{ApplyWindow}(\mathbf{S}_i^{\text{noise}})$ \;
        $\mathbf{A}_w^{\text{init}} \gets \text{ReplaceSegment}(\mathbf{A}_w^{\text{init}}, \mathbf{S}_i, \mathbf{S}_i^{\text{replaced}})$ \;
    }
    
    $\mathbf{A}_w \gets \text{OverlapAdd}(\mathbf{A}_w^{\text{init}})$ \;
}
\Else{
    $\mathbf{A}_w \gets \mathbf{A}_w^{\text{init}}$ \tcp*{No pitch detected, use original}
}

\Return $\mathbf{A}_w$ \;
\end{algorithm}

\section{Code}\label{app:code}
Below are two versions of the code: the original implementation (Original) and the revised version (Modified). After implementing these modifications in the file\footnote{\url{https://github.com/FunAudioLLM/CosyVoice/blob/652132ebaa3133428bf4db7e092a3cd1e073ca80/examples/libritts/cosyvoice3/cosyvoice/flow/flow.py}}, we were able to successfully fine-tune the Flow model within CosyVoice3.
\begin{lstlisting}[language=Python, caption=Original, basicstyle=\ttfamily\small]
def forward(
        self,
        batch: dict,
        device: torch.device,
) -> Dict[str, Optional[torch.Tensor]]:

    # Data extraction
    token = batch['speech_token'].to(device)
    token_len = batch['speech_token_len'].to(device)
    feat = batch['speech_feat'].to(device)
    feat_len = batch['speech_feat_len'].to(device)
    embedding = batch['embedding'].to(device)

    # NOTE unified training, static_chunk_size > 0 or = 0
    streaming = True if random.random() < 0.5 else False

    # xvec projection
    embedding = F.normalize(embedding, dim=1)
    embedding = self.spk_embed_affine_layer(embedding)

    # concat text and prompt_text
    mask = (
        (~make_pad_mask(token_len))
        .float()
        .unsqueeze(-1)
        .to(device)
    )
    
    token = (
        self.input_embedding(torch.clamp(token, min=0))
        * mask
    )

    # text encode
    h, h_lengths = self.encoder(
        token,
        token_len,
        streaming=streaming
    )
    h = self.encoder_proj(h)

    # get conditions
    conds = torch.zeros(feat.shape, device=token.device)
    for i, j in enumerate(feat_len):
        if random.random() < 0.5:
            continue
        index = random.randint(0, int(0.3 * j))
        conds[i, :index] = feat[i, :index]
    conds = conds.transpose(1, 2)

    mask = (
        ~make_pad_mask(
            h_lengths.sum(dim=-1).squeeze(dim=1)
        )
        .to(h)
    )
    loss, _ = self.decoder.compute_loss(
        feat.transpose(1, 2).contiguous(),
        mask.unsqueeze(1),
        h.transpose(1, 2).contiguous(),
        embedding,
        cond=conds,
        streaming=streaming,
    )
    return {'loss': loss}
\end{lstlisting}

\begin{lstlisting}[language=Python, caption=Modified, basicstyle=\ttfamily\small]
self.token_proj = nn.Linear(input_size, input_size)

def forward(
    self,
    batch: dict,
    device: torch.device,
) -> Dict[str, Optional[torch.Tensor]]:

    # 1. Data extraction
    token = batch['speech_token'].to(device)
    token_len = batch['speech_token_len'].to(device)
    feat = batch['speech_feat'].to(device)
    feat_len = batch['speech_feat_len'].to(device)
    embedding = batch['embedding'].to(device)


    streaming = random.random() < 0.5

    # 2. Speaker embedding processing
    embedding = F.normalize(embedding, dim=-1)
    embedding = self.spk_embed_affine_layer(embedding)

    # 3. Token to embedding mapping
    # Ensure token shape is [B, T]
    if token.dim() == 3:
        token = token.squeeze(-1)

    token = token.long()

    # Token mask: [B, T, 1]
    token_mask = (
        (~make_pad_mask(token_len))
        .unsqueeze(-1)
        .to(device)
    )

    # Embedding lookup
    h = self.input_embedding(token) * token_mask
    h_lengths = token_len

    # 4. Token-to-mel alignment
    h = h.repeat_interleave(self.token_mel_ratio, dim=1)
    token_mask = token_mask.repeat_interleave(
        self.token_mel_ratio, 
        dim=1
    )
    h_lengths = h_lengths * self.token_mel_ratio

    # 5. Conditional mel (cond) preparation
    conds = torch.zeros_like(feat)
    for i, j in enumerate(feat_len):
        if random.random() < 0.5:
            continue
        index = random.randint(0, int(0.3 * j))
        conds[i, index] = feat[i, index]

    conds = conds.transpose(1, 2)  # [B, mel_dim, T_mel]

    # 6. Flow mask preparation
    # Flow expects shape [B, T]
    flow_mask = (~make_pad_mask(h_lengths)).to(device) 
    
    # 7. Flow loss computation
    loss, _ = self.decoder.compute_loss(
        feat.transpose(1, 2).contiguous(),
        flow_mask.unsqueeze(1),          
        h.transpose(1, 2).contiguous(),
        embedding,
        cond=conds,
        streaming=streaming,
    )

    return {'loss': loss}
\end{lstlisting}

\section{Subjective Evaluation}\label{app:subjective}

We developed a web-based automated subjective evaluation interface (Figure~\ref{fig:test3}) to collect real-time listening ratings. Twenty audio stimuli (four per proposed method) were pre-loaded into four Latin-square sequences to minimise order and carry-over effects; the sequence was automatically selected according to the participant ID entered at the start of the session. Each trial started with the automatic playback of one stimulus (24 kHz, 16-bit WAV; approximately 12 s). Immediately after playback, a five-point Likert scale (0 = very bad, 5 = excellent) appeared on the screen and the participant had 5 s to tap the desired score on a touch device or click with a mouse. Responses were timestamped and written to a CSV file that was automatically downloaded at the end of the session.
\begin{figure*}[h]
  \centering
  \includegraphics[width=1\textwidth]{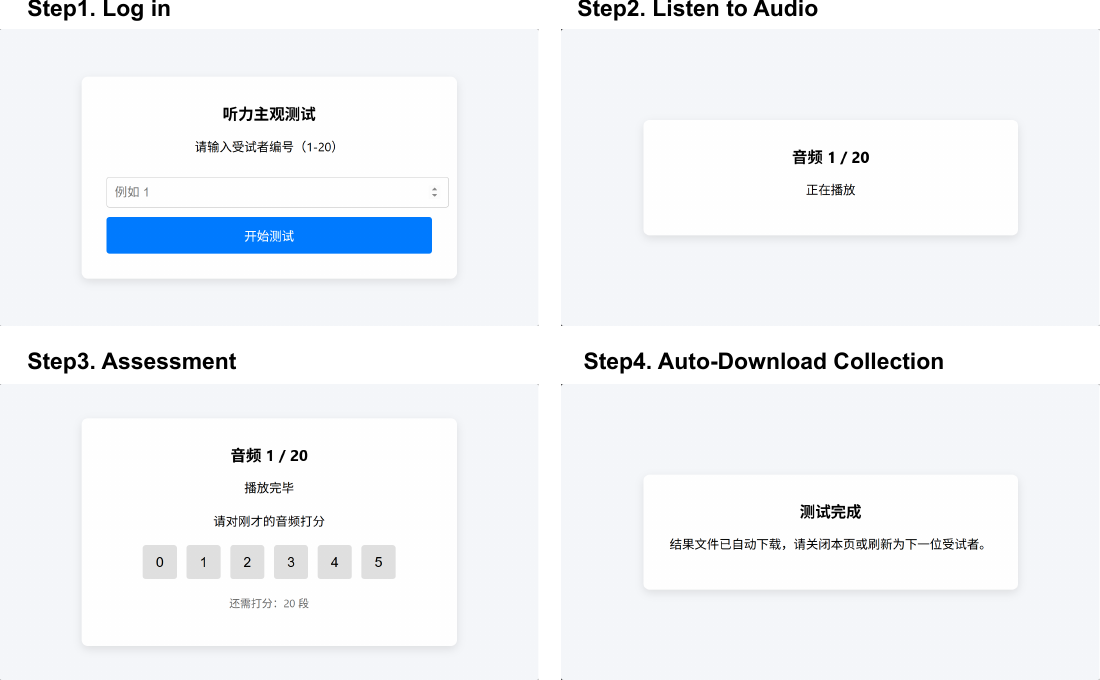}
  \caption{User interface of the listening test.}
  \label{fig:test2}
\end{figure*}

\begin{table*}[h]
  \centering
  \footnotesize
    \begin{tabular}{ c c c c c c c c c } 
    \toprule
        \multirow{2}{*}{\textbf{Language}} & \multirow{2}{*}{\textbf{Method}} & \multicolumn{7}{c}{\textbf{Evaluation Metrics}} \\
        \cmidrule(lr){3-9} 
        ~ & ~ & \textbf{W-MOS} $\uparrow$& \textbf{UTMOS} $\uparrow$& \textbf{DNSMOS} $\uparrow$& \textbf{MCD} $\downarrow$& \textbf{CER} $\downarrow$& \textbf{SpkSim} $\uparrow$& \textbf{VTR} $\downarrow$\\
    \midrule
        \multirow{2}{*}{\textbf{Korean}} 
        & \textbf{CosyVoice3} & 3.35$\pm$1.70 & 1.72 & 2.84 & 62.07 & 27.21 & 0.50 & 0.83 \\
        & \textbf{CosyWhisper (Ours)} & 3.96$\pm$1.04 & 1.49 & 2.84 & 67.67 & 23.32 & 0.53 & 0.75 \\
    \midrule 
        \multirow{2}{*}{\textbf{Japanese}} 
        & \textbf{CosyVoice3} & 2.88$\pm$1.36 & 1.49 & 2.88 & 61.41 & 39.42 & 0.54 & 0.96 \\
        & \textbf{CosyWhisper (Ours)} & 4.03$\pm$0.58 & 1.38 & 2.89 & 69.33 & 36.22 & 0.52 & 0.82 \\
    \bottomrule
    \end{tabular}
  \caption{Evaluation of whisper speech synthesis on Korean and Japanese.}
  \label{tab:multilingual}
\end{table*}

\newpage
\section{Multilingual Performance}
Our CosyWhisper possesses strong inherent potential for multilingual extension because it utilizes the CosyVoice3 tokenizer, which natively supports 9 languages and 18 Chinese dialects.

The limitation of multilingual evaluation stems from a severe lack of publicly available whisper datasets outside of English and Mandarin. Despite this data constraint, To demonstrate this capability, we performed preliminary experiments on Japanese and Korean. We manually curated gender-balanced whispered utterances from YouTube ASMR videos (two male, two female speakers per language), and invited native speakers to evaluate the generated outputs. The results in Table~\ref{tab:multilingual} indicates that our model successfully generalizes and maintains authentic whisper qualities in these new languages. 

To further support our claims, we will add these generated Korean and Japanese whisper samples to our revised supplementary materials, illustrating the generalization capability of our framework while discussing current data constraints.

\section{Case Study}
\begin{figure*}[ht] 
\centering

\begin{minipage}{0.98\linewidth} 
    \small \raggedright
    \textbf{[Transcription]:} "So for a day or two, the Hands refused to pick up food, the Mouth refused to receive it, and the Teeth had no work to do." \\
    \textbf{[Dataset]:} "CHAINs", \textbf{[Sentence ID]:} "f04-03", \textbf{[Speaker]:} "irm06"
\end{minipage}

\vspace{1em}

\begin{subfigure}[b]{0.32\linewidth}
    \centering
    \includegraphics[width=\linewidth]{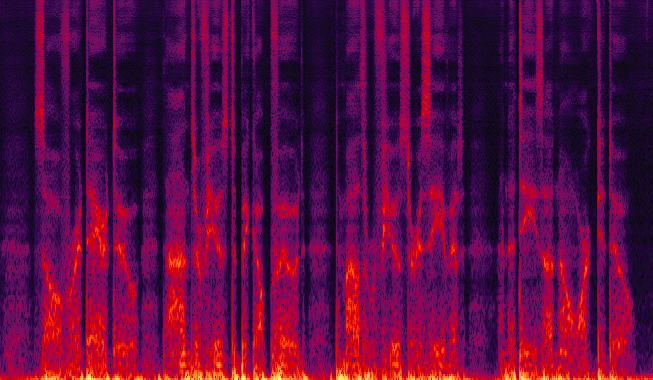}
    \caption{CosyWhisper} 
\end{subfigure}\hfill
\begin{subfigure}[b]{0.32\linewidth}
    \centering
    \includegraphics[width=\linewidth]{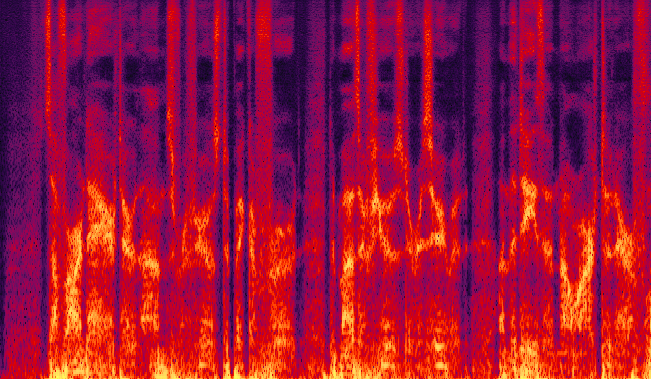}
    \caption{CosyVoice FT}
\end{subfigure}\hfill
\begin{subfigure}[b]{0.32\linewidth}
    \centering
    \includegraphics[width=\linewidth]{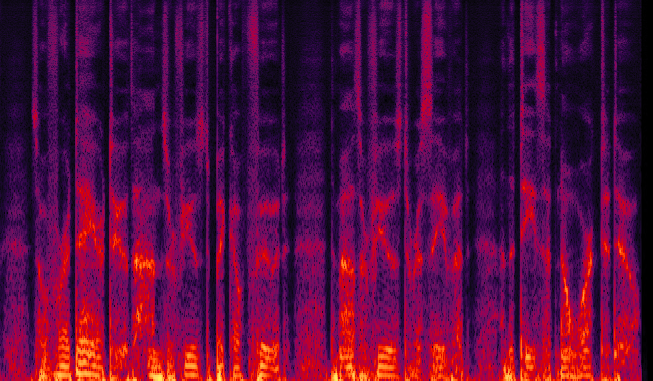}
    \caption{CosyVoice Rep} 
\end{subfigure}

\vspace{1.5em} 

\begin{subfigure}[b]{0.32\linewidth}
    \centering
    \includegraphics[width=\linewidth]{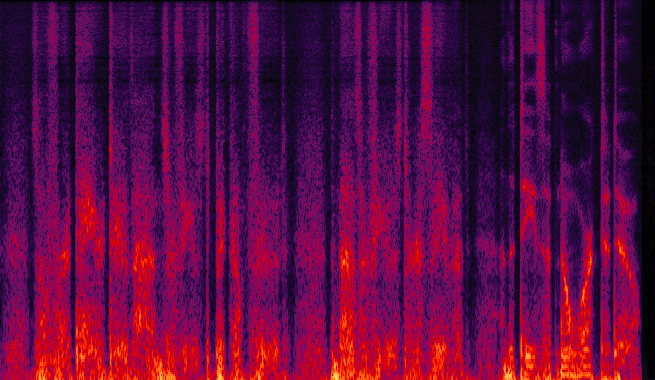}
    \caption{WhispSynth}
\end{subfigure}\hfill
\begin{subfigure}[b]{0.32\linewidth}
    \centering
    \includegraphics[width=\linewidth]{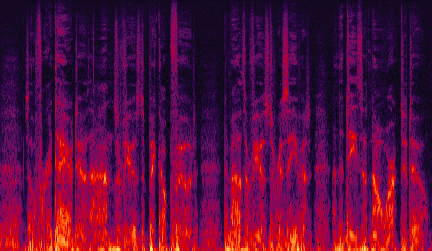}
    \caption{WhispReal}
\end{subfigure}\hfill
\begin{subfigure}[b]{0.32\linewidth}
    \centering
    \includegraphics[width=\linewidth]{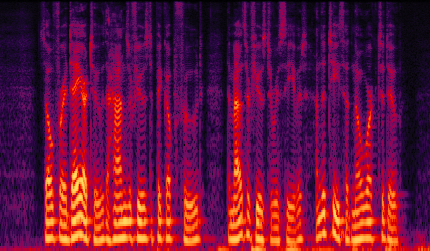} 
    \caption{Normal Speech} 
\end{subfigure}

\vspace{3em} 

\begin{minipage}{0.98\linewidth}
    \small \raggedright
    \textbf{[Transcription]:} \begin{CJK*}{UTF8}{gbsn}“河南巩县百花彩印厂非法印刷西安太阳食品集团商标的包装物，长达一年之久。”\end{CJK*} \\
    \textbf{[Dataset]:} "WhispNJU", \textbf{[Sentence ID]:} "B280", \textbf{[Speaker]:} "B082"
\end{minipage}

\vspace{1em}

\begin{subfigure}[b]{0.32\linewidth}
    \centering
    \includegraphics[width=\linewidth]{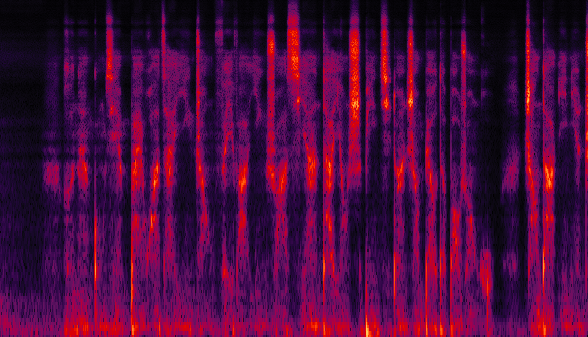}
    \caption{CosyWhisper}
\end{subfigure}\hfill
\begin{subfigure}[b]{0.32\linewidth}
    \centering
    \includegraphics[width=\linewidth]{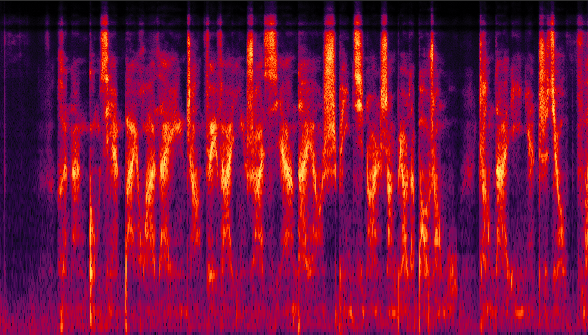}
    \caption{CosyVoice FT}
\end{subfigure}\hfill
\begin{subfigure}[b]{0.32\linewidth}
    \centering
    \includegraphics[width=\linewidth]{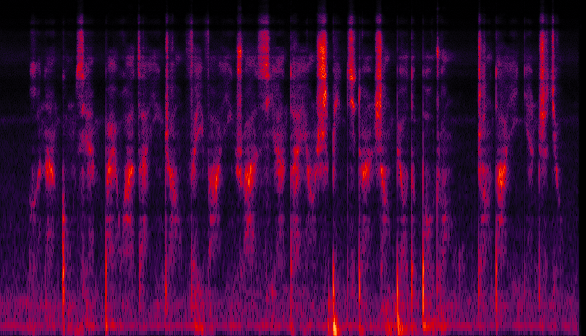}
    \caption{CosyVoice Rep}
\end{subfigure}

\vspace{1.5em}

\begin{subfigure}[b]{0.32\linewidth}
    \centering
    \includegraphics[width=\linewidth]{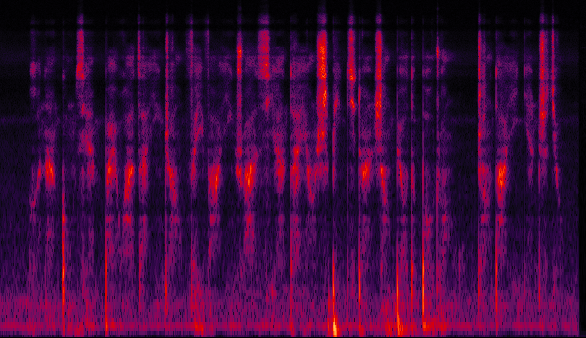}
    \caption{WhispSynth}
\end{subfigure}\hfill
\begin{subfigure}[b]{0.32\linewidth}
    \centering
    \includegraphics[width=\linewidth]{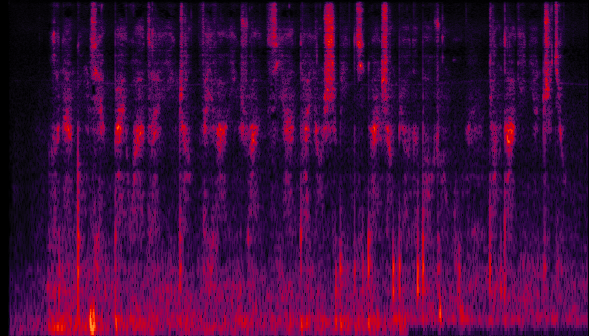}
    \caption{WhispReal}
\end{subfigure}\hfill
\begin{subfigure}[b]{0.32\linewidth}
    \centering
    \includegraphics[width=\linewidth]{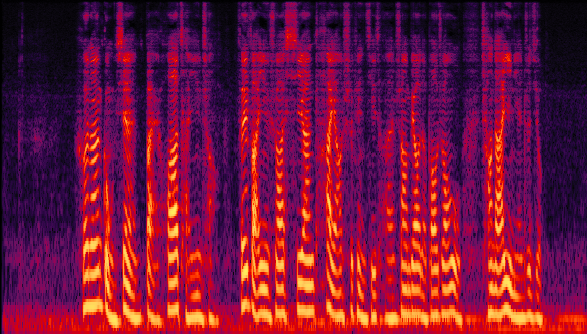} 
    \caption{Normal Speech} 
\end{subfigure}

\caption{Sample spectrograms from two datasets. (a-f) English samples from the CHAINs Dataset. (g-l) Chinese samples from the WhispNJU Dataset. (a,g) CosyWhisper; (b,h) CosyVoice3 Finetuned on WhispReal; (c,i) CosyVoice3 Repeat; (d,j) WhispSynth; (e,k) WhispReal; (f,l) Normal Speech.}
\label{fig:spec_samples}
\end{figure*}

Sample spectrograms are provided in Figure~\ref{fig:spec_samples}; additional examples and the full demo are available on our demo website \url{https://github.com/tan90xx/cosywhisper}.

\end{document}